# IF THE PROSPECT OF SOME OCCUPATIONS ARE STAGNATING WITH TECHNOLOGICAL ADVANCEMENT? A TASK ATTRIBUTE APPROACH TO DETECT EMPLOYMENT VULNERABILITY

**Iftekhairul Islam**

School of Economics, Public Policy, and Political Science
The University of Texas at Dallas
E-mail: iftekhairul.islam@utdallas.edu

**Fahad Shaon**

Erik Jonsson School of Engineering and Computer Science
The University of Texas at Dallas
Email: fahad.shaon@utdallas.edu

**Abstract**

Two distinct trends can prove the existence of technological unemployment in the US. First, there are more open jobs than the number of unemployed persons looking for a job, and second, the shift of the Beveridge curve. There have been many attempts to find the cause of technological unemployment. However, all of these approaches fail when it comes to evaluating the impact of modern technologies on employment future. This study hypothesizes that rather than looking into skill requirement or routine non-routine discrimination of tasks, a holistic approach is required to predict which occupations are going to be vulnerable with the advent of this 4th industrial revolution, i.e., widespread application of AI, ML algorithms, and Robotics. Three critical attributes are considered: bottleneck, hazardous, and routine. Forty-five relevant attributes are chosen from the O*NET database that can define these three types of tasks. Performing Principal Axis Factor Analysis, and K-medoid clustering, the study discovers a list of 367 vulnerable occupations. The study further analyzes the last nine years of national employment data and finds that over the previous four years, the growth of vulnerable occupations is only half than that of non-vulnerable ones despite the long rally of economic expansion.

## 1. Introduction

Unemployment has been a crucial issue for governments as well as policymakers around the world for quite a long time. It has a consequence not only for the social, political and economic life of individuals but also for the overall economy and development of a country (Dickens et al. 1994). The social and political enterprise of people surrounding an unemployed person are also adversely affected. (McClelland et al. 2000). As unemployment has extensive ramifications in different policy areas ranging from welfare, education, tax and investment to health, crime, and other social issues as well as to political stability, both federal and state governments are keen to address this problem with all the tools within their arsenal, i.e., fiscal and monetary policies and other administrative strategies and schemes. However, without an effective and robust prediction mechanism concerning future employment scenarios, such policies would be pointless.

Therefore it is necessary to have a deep understanding of why unemployment occurs in the first place. There could be multifaceted reasons. According to Krugman (1994), the causes of unemployment can be separated into two broad categories: cyclical and structural. The former depends on fluctuations of aggregate demand and the latter on demographic shifts, changes in labor market institutions, or technological innovations.

Unlike cyclical unemployment, which is typically short term, easily predictable, and dependent on market forces, structural unemployment is longer lasting and caused by different externalities which may or may not be directly related to economic indicators (Mocan 1999). Such externalities could be policy shifts, globalization, or new product and process innovations. When there is an initiation of new business or production process, a severe skill mismatch can emerge; companies need competent workers, but the existing workers lack the necessary skills (Manacorda et al. 1999). As it takes substantial time to learn new skills, structural unemployment can last for a long time unless some radical initiatives in training and education are taken.

Lack of skill is not the only reason for structural unemployment. Sometimes a whole region can be affected because of a lack of farsightedness and vision from governments and policymakers. Fagerberg et al. (1997) investigate the causes of structural unemployment in different European countries and find that it prevails more in those countries with less investment in research and development. In these countries the diffusion of technology in production and manufacturing sectors is limited.

We can find a trace of structural unemployment in the US. As of the latest 'Job opening and labor turnout summary' published by BLS in November 2019, the number of job openings is 6.9 million, hires is 5.8 million, and separations (quits, layoffs, and discharges) is 5.3 million. The situation is there is a job for everyone, but still, there is a persistent unemployment rate of 3.8%. Most of these unemployed people are active job seekers but without any success. This scenario is a telltale sign of structural unemployment.

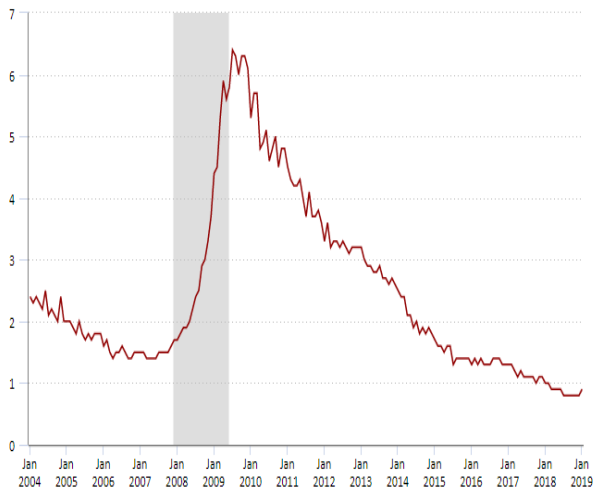

Figure 1: Number of unemployed persons per job opening
Source: US Bureau of Labor Statistics

For the first time since this Job Openings and Labor Turnover Survey (JOLTS) has been published, the ratio has fallen below 1, meaning more number of jobs are offered in the economy than the number of unemployed persons. The shaded area represents recession, as determined by the National Bureau of Economic Research.

There is another way to find out if an economy is facing structural unemployment, the Beveridge curve which shows an inverse relationship between labor demand and labor supply over time. The curve plots job opening rates on the vertical axis and unemployment rate on the horizontal axis. A shift of this curve to the right provides evidence of structural unemployment (Nickell et al. 2001). We can detect an apparent shift of this curve from early 2010 to the end of 2018 comparing to the 2000-2009 period. With further observation, we identify that in the last few months of 2018 the unemployment rate was lower than the job openings rate, confirming the existence of structural unemployment. There is thus a clear indication of skill mismatch between what employers want and what prospective workers can provide.

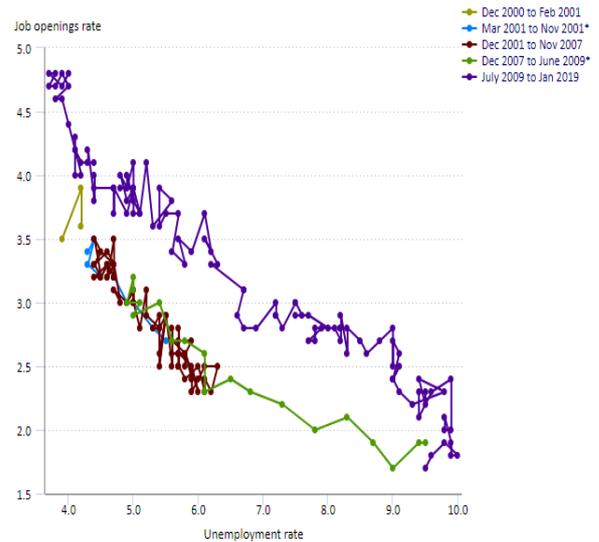

Figure 2: The Beveridge curve
Source: US Bureau of Labor Statistics

## 2. Causes of structural unemployment

There have been many efforts to find the cause of structural unemployment, but none of them has provided conclusions that hold true under every scenario or time frame. The most popular attempt is called the Skill-Biased Technological Change (SBTC) hypothesis, according to which the diffusion of technology increases the demand for high-skilled and educated workers and less-skilled workers become underemployed or sometimes completely out of employment. (Acemoglu 1995). SBTC involves a shift towards capital intensive production techniques that bolstered relative wages of skilled laborers by enhancing relative productivity. This phenomenon was conspicuous in the years after the 2nd World War when the widespread application of technology in production and manufacturing became commonplace (Violante, 2016).

A 'hollowing out of the middle' effect introduced doubts concerning the validity of SBTC theory. The overall demand for jobs is not perfectly linear with skill level and the demand for workers with mid-level skill or mid-ranked jobs has been declining in overall market share since the late 1970s relative to low-skilled and high-skilled workers, as shown by figure 3 and 4.

This phenomenon of job polarization could not be explained by the SBTC hypothesis. To overcome this drawback, Autor et al. (2001) advanced a Routine-Replacing Technological Change (RRTC) or Routine-biased Technological Change (RBTC) hypothesis. As per this theory, job susceptibility due to technological change is not a function of skill. Instead, it is directly related to the types of tasks one performs in doing one's job, as suggested by Table 1.

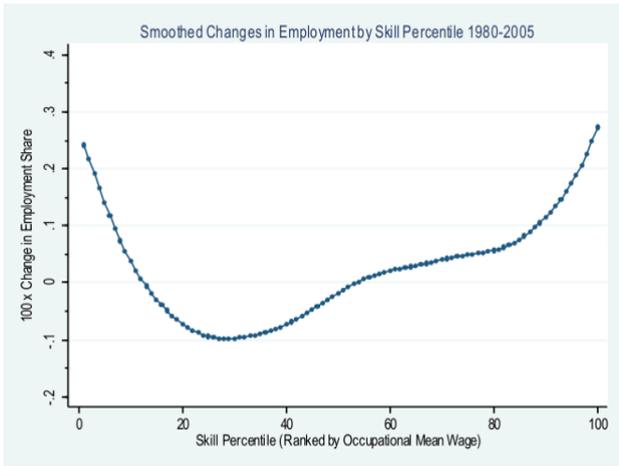

Figure 3: 'Hollowing out of the middle' effect
Source: Autor et al. (2001)

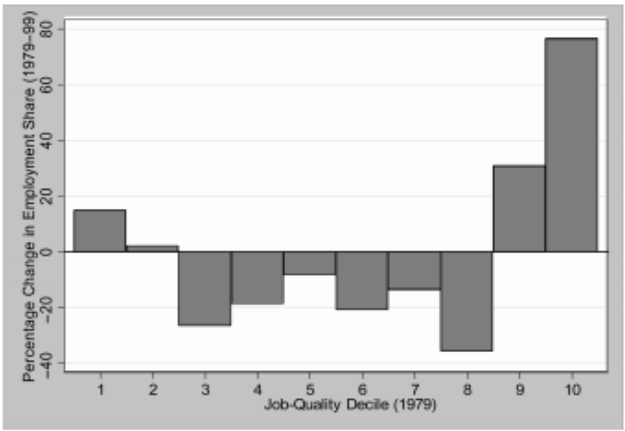

Figure 4: Percentage change in employment share by job-quality decile
Source: Goos et al. (2007)

This phenomenon of job polarization could not be explained by the SBTC hypothesis. To overcome this drawback, Autor et al. (2001) advanced a Routine-Replacing Technological Change (RRTC) or Routine-biased Technological Change (RBTC) hypothesis. As per this theory, job susceptibility due to technological change is not a function of skill. Instead, it is directly related to the types of tasks one performs in doing one's job, as suggested by Table 1.

Table 1: Task metrics of an occupation

| Job type | Manual | Cognitive | |
|---|---|---|---|
| | | Analytical | Interpersonal |
| Routine | Most vulnerable | Moderate Vulnerable | Moderate Vulnerable |
| Non-routine | Moderate Vulnerable | Least Vulnerable | Least Vulnerable |

The hypothesis postulates that routine-manual jobs are the most vulnerable and jobs that demand cognitive input and non-routine maneuvering are the least vulnerable to new and more advanced technology in the production and service sectors. Figure 5 supports of their claim.

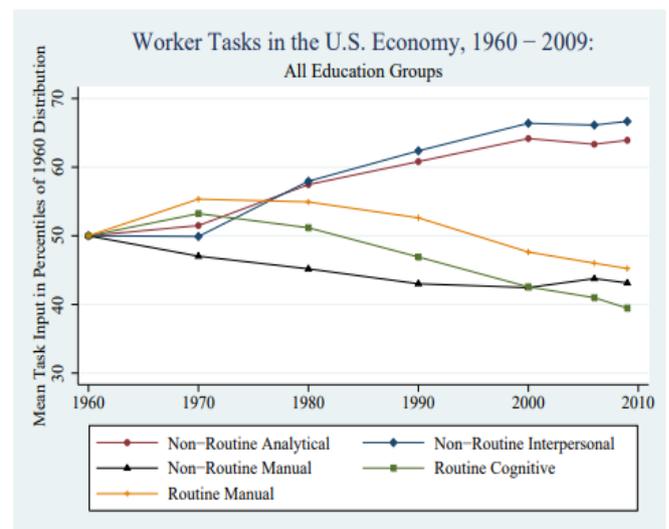

Figure 5: Worker task constituents in the US economy from 1960-2009.
Source: Autor et al., (2013)

While Autor's hypothesis explain why the number of middle-ranked jobs is declining they have become irrelevant with the advent of Artificial Intelligence (AI), Machine Learning Algorithms, and high-end, agile robots. The tasks the new machines and algorithms perform cannot be defined by limited routine, non-routine or cognitive, manual task metrics. To predict job vulnerability due to this latest technology, it is necessary to take into consideration not only the skill level of workers or task constituents of a job, but all job requirement variables, i.e., abilities, capacities, experience, training and education, job context, etc. (Frey et al., 2017). A holistic approach has to be taken to grasp the true impact of the latest trends in technology on the overall job market.

There has been no recent attempt to pinpoint which jobs are susceptible and which are not, taking into account this very recent trends in innovation i.e., AI and Robotics. The previous literature has tried to detect vulnerability by sector (manufacturing, service, agriculture), or industries (health, education, finance, management, etc.) or types of jobs (routine or non-routine), or skill requirement of jobs (high or low)

We know of no undertaking that investigates why the prospect of some employments are diminishing over the past several years despite the long rally of economic expansion. This study would like to shed some light on this issue in the line with 'task- attribute' approach, which is an amalgamation (broadly) of the classical hypotheses and present-day analysis and findings.

## 3. Material and methods

### 3.1. Datasets

For the study, two sources of datasets have been used, Bureau of Labor Statistics (BLS) and Occupational Information Network (O*NET). BLS provides full spectrum employment data from all the USA states and its territories. It also categorizes the data in terms of sectors and industries. This study has used employment data from 2010 to 2018. It is only concerned with the total number of national-level employment from each SOC (Standardized Occupational Code) and how its distribution changed each year. O*NET contains detailed occupational definitions of all the SOC-coded employments of BLS. O*NET explains all the jobs using more than 180 job features, under the broad categories of abilities, context, skills, knowledge, interests, work style, and work activities. The defining process is extremely elaborative and exhaustive. For each job and corresponding features, O*NET provides with an importance level of the feature for that particular job. In Figure 6, the study shows a standardized importance score of selected features for some occupations.

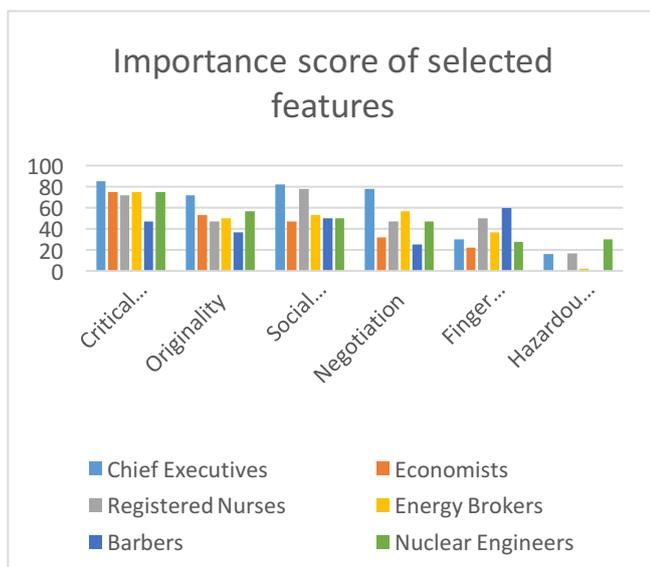

Figure 6: Importance score of selected features

The researchers interview people from different professions and ask them questions, both general and specific to their respective professions. Then they create a Likert scale ranging from 0 to 5. A score of 0 means that attribute is not at all important for that occupation and a score of 5 means very important. Based on these surveys they come up with an importance level for a particular attribute of a particular occupation. After that, the scores are standardized following a general formula. For example, we take the attribute 'speaking' for a lawyer and a paralegal. For both the professions, it is a very important requirement. But, their importance level is different. For lawyer, it is 70 and for a paralegal, it is 50. As these interviews are taken every year, any change in the importance level of an attribute of a particular occupation is reflected in the database within a year.

Cleaning and processing of datasets are imperative steps to make sense of these data. The way O*NET stores the information is difficult for comparison. The importance level of all the attributes (more than 180 variables for most of the occupations) is scattered over 1,110 files. PYTHON scripting facilitates downloading and merging these files. Because not all of the attributes are applicable to AI and robots, a subset of the whole dataset is created. As each attribute has an individual ID, it is used as the primary key to create a relational database using MYSQL.

For this analysis, all the occupations that are present in both O*NET and BLS databases and with relevant SOC codes are considered. After leaving the occupations with missing data, the study is left with 966 occupations. These occupations are the observations for this study.

### 3.2. Variables selection

Most of the relevant variables are initially chosen from the 180 odd variables listed in occupational characteristics based on the 'engineering bottleneck' concept. According to this concept even though AI and robots can perform different non-routine cognitive and non-routine manual tasks, contemporary technology still falls short of expectation when it comes to such tasks as perception and manipulation, creative intelligence, social intelligence, etc. (Frey et al. 2013). The basic premise of the hypothesis is that there are some choke points that cannot be resolved by the current level of technological development when it comes to imitating human behavior, skill, and intuition that are critical for performing some occupational tasks, i.e., judgment, coordination, social perceptiveness, teamwork, etc. On top of these 'bottleneck' variables, we take into account contemporary research and articles on AI and robotics to decide on the occupational attributes that are going to be or already being emulated by AI or robots with higher precision and effectiveness, i.e., pattern recognition, repetitive motion, manual dexterity, etc. These attributes fall into routine work category, i.e., not much critical thinking, complex problem solving, or originality is required, rather following a well-defined set of rules is enough to perform the job properly. The database provides 41 attributes, which are either very difficult to emulate (engineering bottlenecks) or already being in the process of replacement (routine and repetitive) by AI and robots.

There are some contextual attributes of some professions that could be detrimental for human lives, i.e., working in hazardous conditions, exposure to contamination and radiation, etc. As no one can take chances with human lives, these kinds of jobs are sure to be replaced by robots as soon as technology reaches the threshold level. (Takayama et al. 2008), Maney (2018), Clay (2014). For this reason, we have introduced 4 more variables that are related to hazardous and risky tasks. In the end, we have 45 relevant variables to further our analysis of understanding vulnerable occupations due to AI and robots. Table 6, gives an example of some variables as per these three types of broad features. Finally, our database now constitutes a 966X45 matrix reduced from the initial 1110X180 matrix due to data availability and relevancy issue.

Table 6: Features and contexts

| Relevant Features | Selected variables |
|---|---|
| Engineering bottleneck | Negotiation, Critical thinking, Empathy, Persuasion etc. |
| Hazardous context | Exposed to Contaminants, Very Hot or Cold Temperatures etc. |
| Routine work | Finger Dexterity, Repetitive motions etc. |

### 3.3. Data reduction using Principal Axis Factor Analysis

After merging and cleaning the dataset, 966 occupations remain, along with the 45 necessary and relevant variables scaled by their respective importance levels. As many of these variables were correlated (both Barlett sphericity and Kyser-Meyer-Olkin tests are performed), a principal axis factor analysis (PAFA) is undertaken to isolate the independent dimensions of variation. Principal axis factor analysis is a statistical technique that helps to reduce a large set of independent variables to a smaller and meaningful set of summary variables. It assists in exploring the latent construct of a phenomenon that may be present in several independent variables simultaneously.

#### 3.3.1. Number of factors

Before performing the factor analysis, the study has performed 'Parallel Analysis' to find the optimum number of factors. In this case, the number could be from 6 to 8.

After performing the analysis, the best result is 7 factors, a solution that has the highest Tucker-Lewis value of 0.83. The root mean square of residuals (RMSR) is 0.02. This is acceptable as this value should be close to 0. The RMSEA index is 0.092, and BIC is 1477.06. Both these values are better than factor results with other factor numbers.

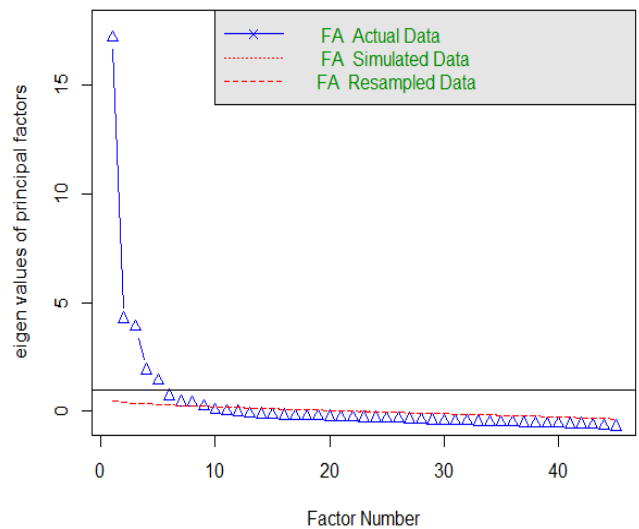

Figure 7: Parallel Analysis scree plot

#### 3.3.2. Factor loadings

Figure 8 shows how the original variables loaded onto each factor in the 7-factor orthogonally rotated PAFA solution.

#### 3.3.3. Interpretation of factors

We interpret the six basic skill-dimensions as follows:

*Problem-solving and repetitive work*
At one end, this factor involves spotting complex problems and assessing relevant information to develop and evaluate different options and implement the best solution. It also requires the performance of cost-benefit analysis of potential actions using logic and reasoning. This factor is important for identifying the strengths and weaknesses of alternative solutions, conclusions or approaches to problems. At the opposite end of the scale is repetitive work, doing a 'patterned' job that doesn't require out of the box thinking. Such jobs can be very easily codified, as there is not much dispersion from the linear and well-defined job description. As per the Occupational Safety and Health Administration (OSHA), "a highly repetitive job can be characterized by one of the following: A cycle time less than 30 seconds. Over 1,000 parts per shift, or more than 50% of the cycle time involving the same kind of fundamental cycle". Repetitive work can be harmful to a worker as it could be instrumental to musculoskeletal disorders. (PSHSA, 2010)

*Negotiation and leadership capacity*
This factor involves getting others onto the same page and trying to bridge the gaps in opinions and attitudes. It also requires adjusting one's actions in tandem to the counterpart's actions. Sometimes it could also demand to persuade others to change action or behavior as per the changing contexts. This factor is also important in handling and resolving complaints, discontents, and grievances.

*Exposure to hazard*

This factor requires doing a job that is exposed to contaminants (such as pollutants, gases, dust or odors), radiation (gamma, X-rays), risky machineries (moving parts, overhead cranes), very hot (above 90 F degrees) or very cold (below 32 F degrees) temperatures, loud noise (construction sites). Some jobs involve working in a high, uneven places (tree pruning, roofing). Few professions i.e., soldiers, deep-sea explorers, and firefighters are inherently exposed to various vulnerable scenarios. All such jobs that are exposed to hazardous conditions require a high level of depth perception (distinguishing between near and distant objects) and manual dexterity (agile body movement).

Empathy
This factor means being conscious of others' reactions and grasping why they react this way. It also requires being sensitive to others' needs and accommodating any legitimate demand. This factor involves being amiable to others on the job and demonstrating a cooperative, gracious attitude. It entails deep comprehension of human behavior and conduct and individual differences in ability, performance, personality, and interests.

Artistic ability
This factor involves researching and developing new applications, processes or products. Sometimes it also requires the worker to design, create and implement unique, state-of-the-art ideas and schemes.

Coordination/Leading capacity
This factor entails engaging everyone to pursue a common goal. Coordination involves funneling and optimizing everyone's effort to achieve the target of the team as a whole. This factor helps in creating synergy among all the team players so that they can deliver what is expected from them. Knowledge over other team member's limitations and strengths is also a key component of this factor.

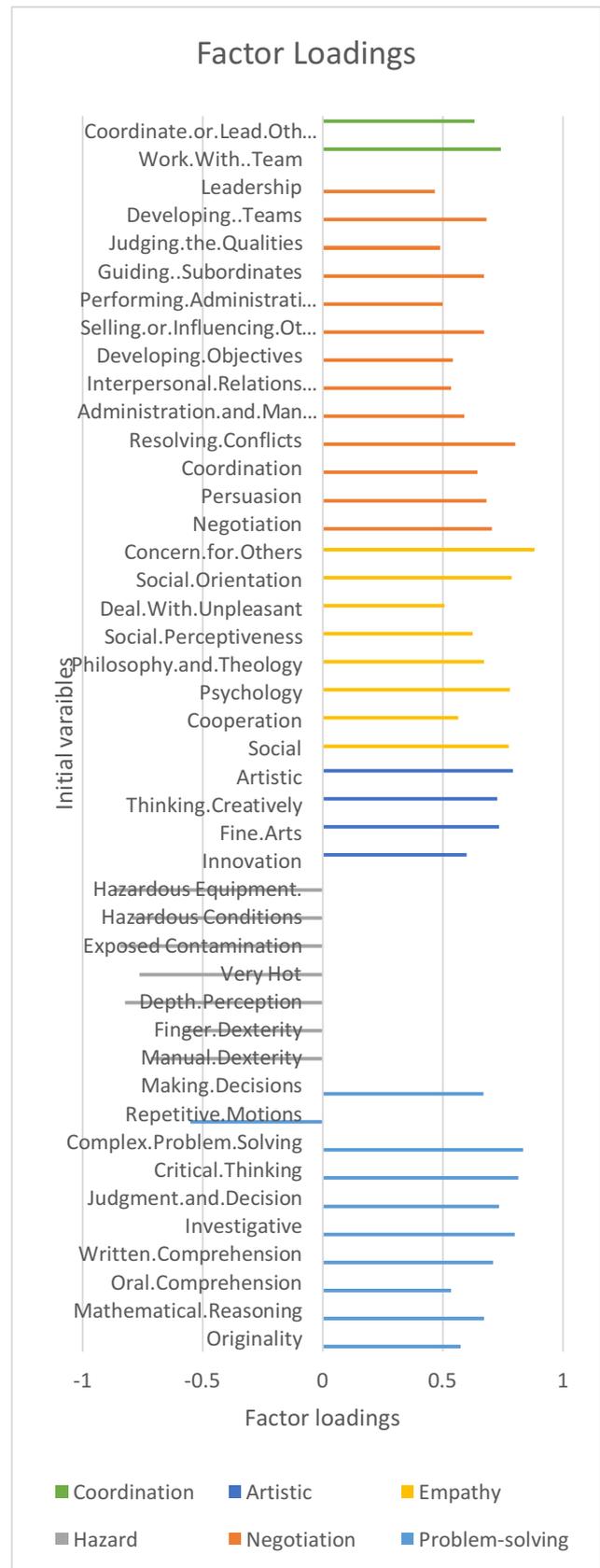

Figure 8: Factor loadings

## 3.4 Susceptibility to AI and Robotics

If we observe closely we see that the factor 'Hazard' i.e., risk exposure (Ford 2015), and dexterity (Strickland 2016) is the most susceptible characteristic in terms of the scope and capacity of modern AI algorithms and robotics. Factor 'problem solving' displays an inverse relation between problem-solving and robotics- susceptible repetitive motion. Occupations with low or negative scores on this factor are vulnerable. Other factors i.e., negotiation capacity, empathy, coordination and artistic are not vulnerable. (Russel et al. 2016). They are extremely difficult for AI and robots to emulate with the current level of research and development. This study then focuses on these susceptible factors to derive the list of occupations that are most likely to be affected by contemporary technological innovation. The list is created by taking the top 20 percent of occupations that scored high on the 'Hazard' factor and the bottom 20 percent of occupations that scored low on the 'Problem-solving' factor. Seven types of susceptibility are found based on these factors. (Figure 9)

## 3.5 Conditioned Vulnerability: K-Medoid Clustering

Not all occupations with similar vulnerabilities are equally at risk from technological change. Vulnerabilities need to be adjusted by accounting for other factors. For example, we can take the cases of barbers and surgeons. Both these professions are heavily dependent on dexterity with factor scores of 2.85 and 1.55 respectively. These occupations might be at risk due to cutting-edge development in robotic
arms. But to perform the job of a surgeon unlike a barber, other factors are also critical. If we take into account other occupational attributes (critical thinking, coordination, judgment and decision, empathy), we can understand how distinct these two professions are in terms of replicability. We, therefore, need an adjusted classification of occupations that merges vulnerability with conditioning characteristics. For that we must take into consideration the 'bottleneck' attributes that are difficult to emulate.
To that end, we use k-medoid clustering (PAM, as Partitioning around Medoids), in the full 7-factor space that took into consideration all the relevant factors, vulnerable and 'anti' vulnerable.

### 3.5.1. Optimum Number of Clusters
To proceed with the k-medoid clustering procedure, we have to first decide on the number of optimum clusters. The fundamental characteristics of any clustering algorithm are that distance (similarity) of elements of the same cluster should be minimum and the intra-cluster distance (dissimilarity) should be maximum. (Rousseeuw P. J., 1987). The 'Silhouette Coefficient' technique is used to derive an initial number of clusters in the 6-space. Silhouette, s(i) is calculated using the following function:

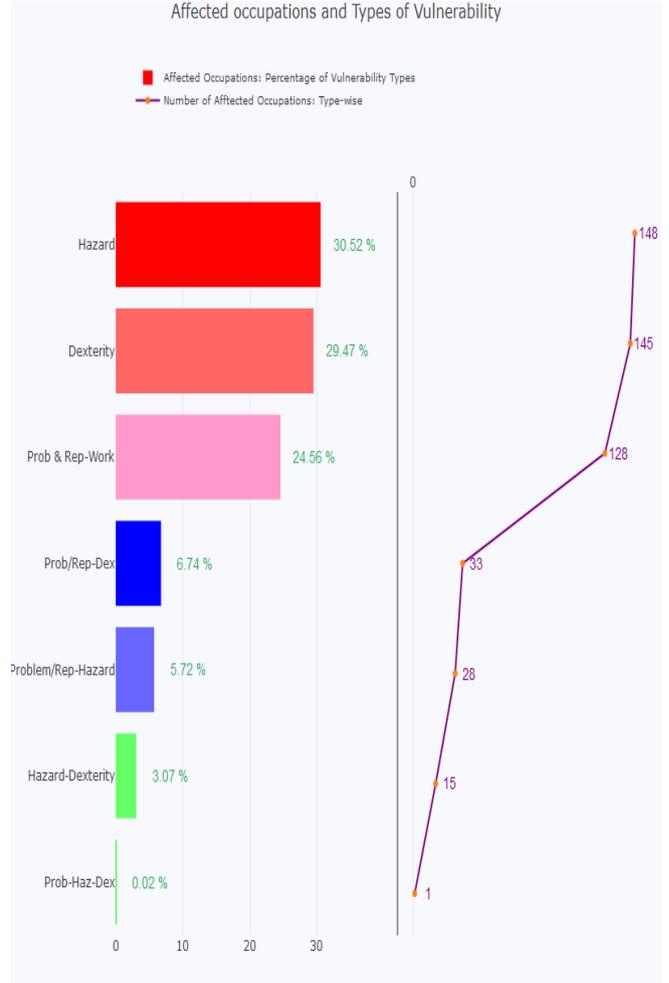

Figure 9: Percentage of affected occupations based on factor-score

$$s = \begin{cases} 1-a(i)/b(i) & if\ a(i) < b(i), \\ 0 & if\ a(i) = b(i), \\ b(i)/a(i) - 1 & if\ a(i) > b(i), \end{cases}$$

where,

$a(i)$ is average dissimilarity of an element i to all other elements within a cluster,

$d(I,C)$ is average dissimilarity of I to all the elements of another cluster $C$,

$b(i)$ is minimum $d(i,C)$

The above function can be rewritten as the following

$$s(i) = \frac{b(i) - a(i)}{max\{a(i), b(i)\}}$$

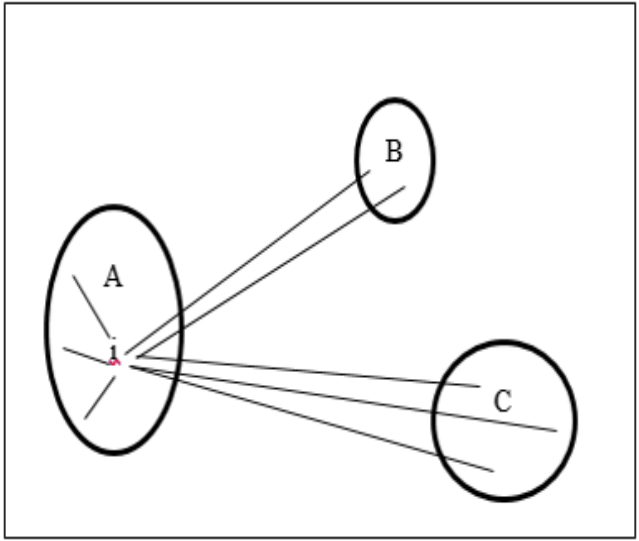

Figure 10: An illustration of the computation of silhouette distance *s(i)*

We can see silhouette value is highest for 7 number of clusters. If we take more than or less than 7 clusters silhouette value decreases. (Figure 11)

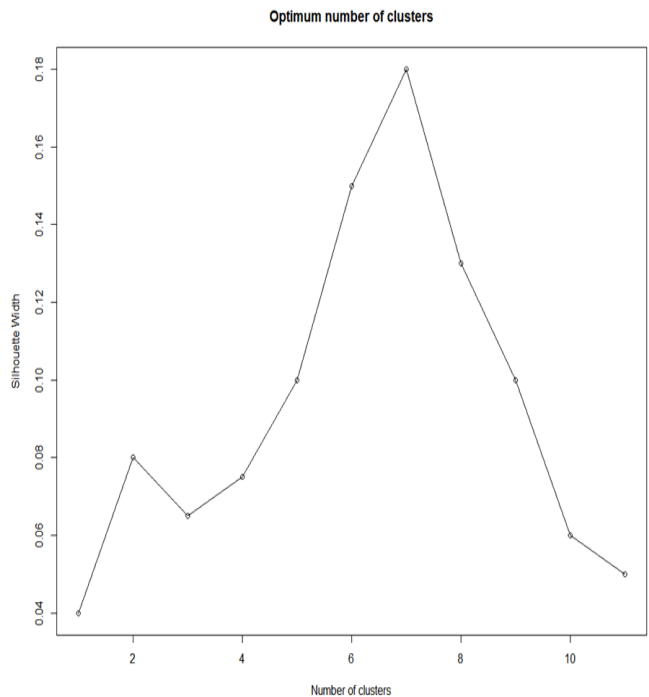

Figure 11: Optimum number of clusters

Figure 13 is a silhouette of the 7 clusters in the 7-space, with medoid results and sample occupations detailed in the appendix. Of the 8 clusters of occupations, only 3 have an immediate vulnerability (3, 6 and 7), located on one side of the red 'vulnerability bisector' in the figure.

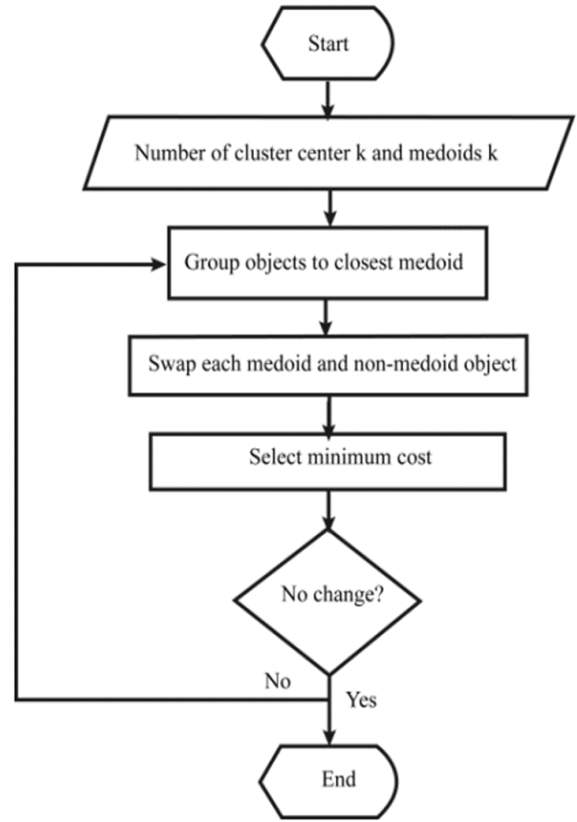

Figure 12: K-medoid clustering method
Source: Pramudita et al. (2018)

### 3.5.2. K-Medoid Clustering

This algorithm (known as Partitioning around medoid) was first proposed in 1987 by two famous mathematicians, Kaufman and Rousseeuw. It is more robust to outliers and noises (Han et al. 2011) than the more popular clustering algorithm k-means clustering. While k-means uses a mean point as the center of the cluster (which may not be a real point in that cluster), k-medoid uses an actual point (medoid) or member in the cluster to categorize it. A medoid of a particular cluster is the most representative element of the cluster, i.e., a medoid's similarities with the other member of the same cluster is the minimum.

### 3.5.3. Procedure for K-Medoid Clustering

K-medoid initiates with selecting a random element for each number of cluster, i.e. k number of medoids for k number of clusters. All other elements are included in the nearby cluster in terms of the features or factor scores for the current scenario. After the initial point, a new medoid is chosen for every cluster and it is tested against the current medoid to see if the value cost function (dissimilarities between each data item and its corresponding medoid) is less or more. If cost is less than the previous medoid than the new point becomes the medoid for that cluster. After a sufficient number of iterations, the algorithm converges

$$Z = \sum_{i=1}^{k} \sum |x - m_i|$$

Z: Sum of error (absolute) for all the points of the dataset
x: A data point in the data space
$m_i$ : Medoid for cluster $C_i$

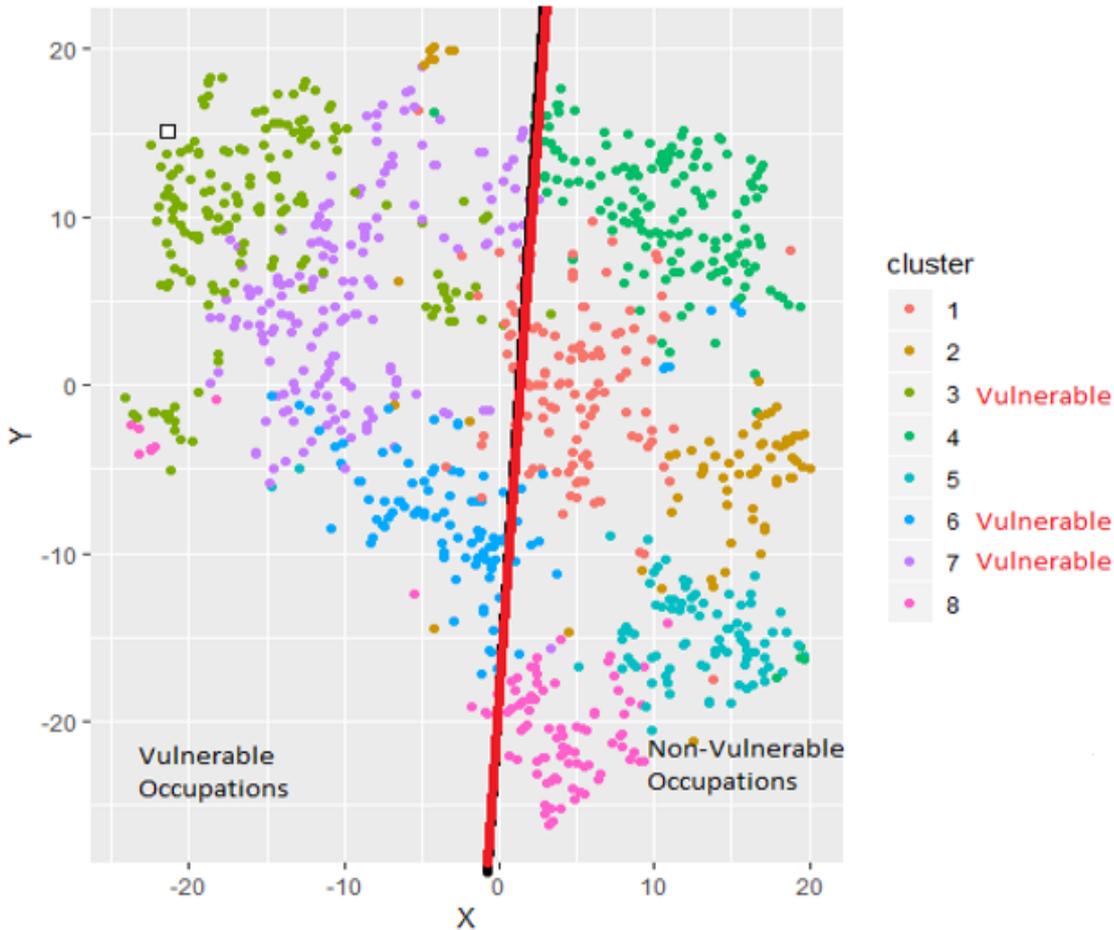

Figure 13: PAM clustering of occupations

### 3.5.4. K-Medoid Result

Of the 8 clusters of occupations, we can find only 3 of them have an immediate vulnerability because these clusters have high factor scores on susceptible factors or very low scores on 'bottle-neck' factors. (Figure 14). On the contrary, 5 clusters have high factor scores in more than 2 'bottle-neck' factors making them not immediately vulnerable. (Figure 15). There are altogether 408 occupations in these 3 clusters. A list of the occupations is provided in the appendix. One thing we must understand, here no attempt is made to make a ranking of vulnerability, i.e., high-mid-low vulnerability. Rather the study just looks at the vulnerable factors and segregates the occupations in terms of factor scores using k-medoid clustering algorithms.

A cursory look of the occupations gives us the idea that most of the jobs are low paid, highly repetitive, low skilled, and doesn't require more than a high school degree. Most of the occupations are from mining, agriculture, transportation, retail, and manufacturing industries. On the other hand, not surprisingly health care, hospitality, and education industries seem to be immune to these trends of prolific application of AI, and Robotics in workplace. Professional and business industries and state and federal government employees are also not so vulnerable.

### 3.6 Real Job Trends

The study further looks at the change of a number of jobs from all the occupations mentioned both in O*NET and BLS for the past 8 years. Initially, I measured the change in demand for jobs for all occupations. The study found that on average, the increase in jobs for the vulnerable occupations was only a little over 1%, whereas the increase in demand for the non-vulnerable jobs was almost 2.4% in these years. When I looked at this trend further, I noticed something interesting. I detected that the number of jobs that were more affected by AI and ML algorithms was decreasing at a higher rate (1.8% precisely). But, many of the vulnerable occupations that would mostly depend on dexterity or other 'Robotic' input were increasing, i.e., truck driver, firefighter etc. Again, not surprisingly number of warehouse helpers decreased two folds during the same period. My conclusion was that, even though current trends in robotics might put many jobs in a vulnerable condition, the threshold point of technology to replace human labor is still far away, and as of now, robots can replace them only in a 'controlled' environment.

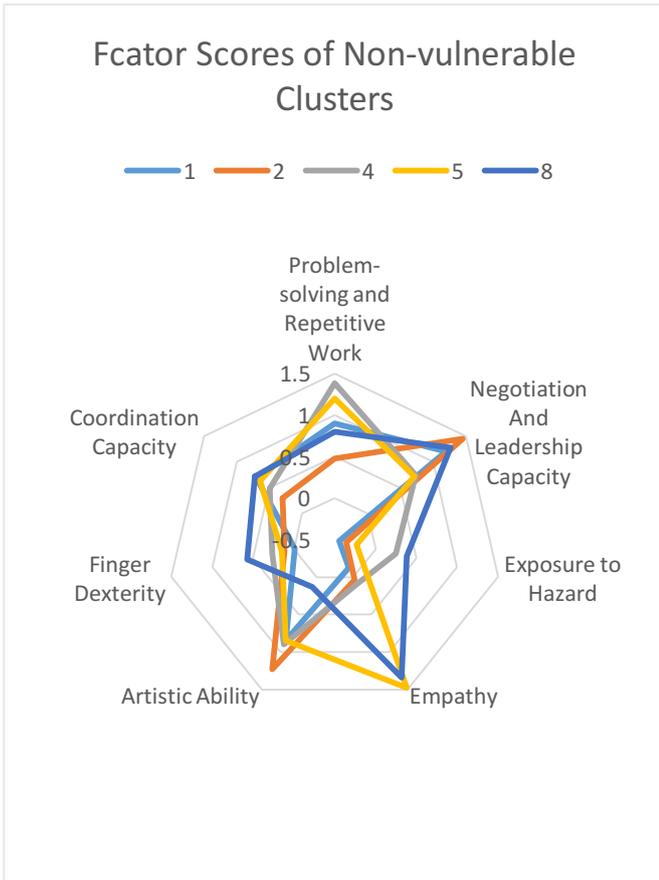

Figure 14: Factor scores of non-vulnerable clusters

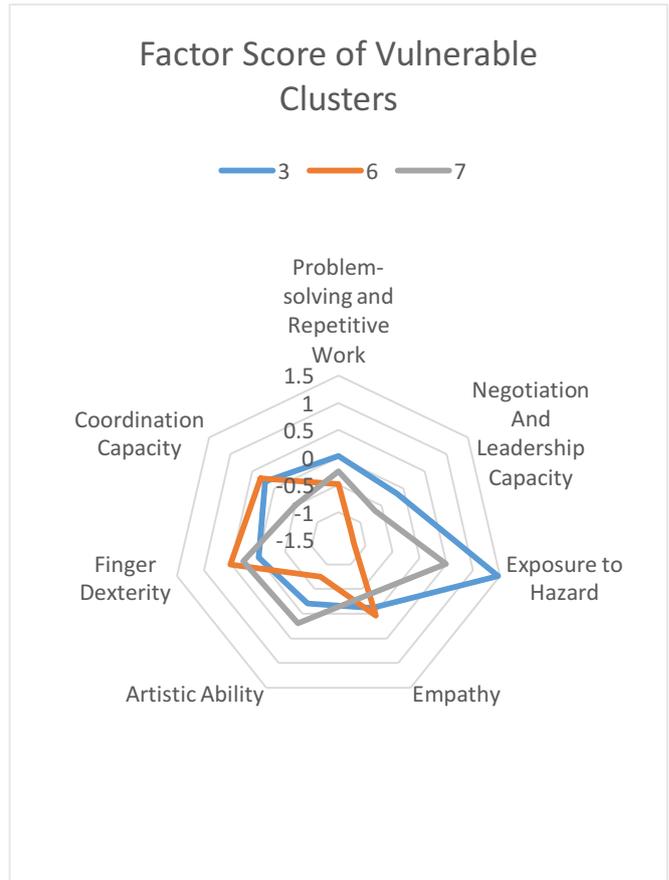

Figure 15: Factor scores of vulnerable clusters

## 4. Conclusion

We can see the demand for quite a handful of occupations is decreasing over the past several years. As, the US economy has not been experiencing an economic downturn, or no major trade agreement or disruptions has taken place during these years (trade war with China is rather a contemporary issue and any major impact is yet to be seen), we can conclude this phenomenon is due to the technological shift in businesses and manufacturing sectors. We must figure out what occupations will be in demand and what relevant skills are required to fill up the positions. Otherwise, the US economy could hit a sudden roadblock, which could end up in a catastrophe.

### References

1. Acemoglu, D. (1995). Reward structures and the allocation of talent. European Economic Review, 39(1), 17-33. doi:10.1016/0014-2921(94)00014-q
2. Acemoglu, D. (2002). Technology and the labor market. Journal of Economic Literature 40, 7–72.
3. Acemoglu, D., & Autor, D. (2010). Skills, Tasks and Technologies: Implications for Employment and Earnings. doi:10.3386/w16082
4. Acemoglu, D., & Restrepo, P. (2016). The Race Between Machine and Man: Implications of Technology for Growth, Factor Shares and Employment. doi:10.3386/w22252
5. Acemoglu, D., & Restrepo, P. (2018). Artificial Intelligence, Automation and Work. doi:10.3386/w24196
6. Arntz, M., T. Gregory and U. Zierahn (2016), The Risk of Automation for Jobs in OECD Countries: A

Comparative Analysis. OECD Social, Employment and Migration Working Papers, No. 189, OECD Publishing, Paris, https://doi.org/10.1787/5jlz9h56dvq7-en.
7. Asher, R. (1995). Work Skill in Historical Perspective. The New Modern Times: Factors Reshaping the World of Work, ed. DB Bills, 51-80.
8. Autor, D. (2013). The Task Approach to Labor Markets: An Overview. SSRN Electronic Journal. doi:10.2139/ssrn.2211349
9. Autor, D. (2013). The Task Approach to Labor Markets: An Overview. SSRN Electronic Journal. doi:10.2139/ssrn.2211349
10. Autor, D., Levy, F., & Murnane, R. (2001). The Skill Content of Recent Technological Change: An Empirical Exploration. doi:10.3386/w8337
11. Bekar, C., Carlaw, K. & Lipsey, R. J Evol Econ (2018) 28:1005 https://doi.org/10.1007/s00191-017-0546-0
12. Benzell, S., Kotlikoff, L., Lagarda, G., and Sachs, J. (2015). Robots Are Us: Some Economics of Human Replacement. doi:10.3386/w20941
13. Berg, M. (1980). The machinery Question and the Making of Political Economy 1815-1848, Cambridge: Cambridge University Press.
14. Berger, T., Chen, C., and Frey, B. (2018). Drivers of disruption? Estimating the Uber effect. European Economic Review, 110, 197-210. doi:10.1016/j.euroecorev.2018.05.006
15. Bessen, J. (2015). How Computer Automation Affects Occupations: Technology, Jobs, and Skills. SSRN Electronic Journal. doi:10.2139/ssrn.2690435
16. Brand J. E. (2015). The Far-Reaching Impact of Job Loss and Unemployment. Annual review of sociology, 41, 359–375. doi:10.1146/annurev-soc-071913-043237
17. Bruno, F., Falzoni, A., and Helg, R.(2003). Measuring the effect of globalization on labour demand elasticity: An empirical application to OECD countries. Discussion Paper
18. César, A., and Dolores, C. (2002). Innovation and Job Creation and Destruction . Evidence from Spain. Recherches économiques de Louvain. 68. 148-168. 10.3917/rel.681.0148.
19. Christensen, M. (1997). The Innovator's Dilemma: When New Technologies Cause Great Firms to Fail. University of Illinois at Urbana-Champaign's Academy for Entrepreneurial Leadership Historical Research Reference in Entrepreneurship. Available at SSRN: https://ssrn.com/abstract=1496206
20. Clay, K. (2014). Robots in Risky Jobs: On the Battlefield and Beyond. ISBN 978-1-4765-3972-0. Publisher Capstone Press .
21. Cortes, G. M., Jaimovich, N., and Siu, H. E. (2017). Disappearing routine jobs: Who, how, and why? Journal of Monetary Economics, 91:69–87.
22. Daniel, G. (2016). U.S. Farms Still Feed the World, But Farm Jobs Dwindle. https://www.strategy-business.com/blog/US-Farms-Still-Feed-the-World-But-Farm-Jobs-Dwindle?gko=928b6
23. David, B. (2017). Computer technology and probable job destructions in Japan: An evaluation. Journal of the Japanese and International Economies, 43, 77-87. doi:10.1016/j.jjie.2017.01.001
24. Davidson, P. (2013) Income inequality and hollowing out the middle class, Journal of Post Keynesian Economics, 36:2, 381-384, DOI: 10.2753/PKE0160-3477360209
25. Dickens, R., Machin, S., & Manning, A. (1994). The Effects of Minimum Wages on Employment: Theory and Evidence from the US. doi:10.3386/w4742
26. Edquist, C., Hommen, L., and Mckelvey, M. (2001). Innovation and Employment: Process Versus Product Innovation. 10.4337/9781843762874.
27. Feldmann, H. (2013), Technological unemployment in industrial countries, Journal of Evolutionary Economics, 23, issue 5, p. 1099-1126 https://EconPapers.repec.org/RePEc:spr:joevec:v:23:y:2013:i:5:p:1099-1126.
28. Ford, M. (2009). The Lights in the Tunnel: Automation, Accelerating Technology, and the Economy of the Future, 2009.
29. Ford, M., (2015). Rise of the Robots: Technology and the Threat of a Jobless Future, Basic Books.
30. Frey, C., & Osborne, M. (2017). The future of employment: How susceptible are jobs to computerisation? Technological Forecasting and Social Change, 114, 254-280. doi:10.1016/j.techfore.2016.08.019
31. Galor, O., and O. Moav. (2000). "Ability Biased Technological Transition, Wage Inequality Within and Across Groups, and Economic Growth," Quarterly Journal of Economics 1 15, 469-497.
32. Garud, R., Praveen, N., and Zur, S. (1997). Technological Innovation: Oversights andForesights. New York: Cambridge University Press. ISBN 0-521-55299-0.
33. Georg, G. and Guy, M. (2015). Robots at Work. London School of Economics & Political Science (LSE). IZA Discussion Paper No. 8938
34. Goldin, C., & Katz, L. (2007). The Race between Education and Technology: The Evolution of U.S. Educational Wage Differentials, 1890 to 2005. doi:10.3386/w12984
35. Goos, M., and Manning, A. (2007). Lousy and Lovely Jobs: The Rising Polarization of Work in Britain. Review of Economics and Statistics, 89(1), 118-133
36. Goos, M., Konings, J., and Vandeweyer, M. (2015). Employment Growth in Europe: The Roles of Innovation, Local Job Multipliers and Institutions.

https://ssrn.com/abstract=2671765 or http://dx.doi.org/10.2139/ssrn.2671765
37. Henderson J.P. (1992) Introduction to Thomas Robert Malthus, Principals of Political Economy, 1820. In: Essays in the History of Mainstream Political Economy. Palgrave Macmillan, London
38. Machin, S., and Van, J. (1998), Technology and Changes in Skill Structure: Evidence from Seven OECD Countries, The Quarterly Journal of Economics, 113, issue 4, p. 1215-1244.
39. Manacorda, M., Manning, A., and Wadsworth, J. (2006), The Impact of Immigration on the Structure of Male Wages: Theory and Evidence from Britain, CReAM Discussion Paper No. 08/06.
40. Maney, K. (2019). The Upside to Robots Stealing Jobs? They'll Take the Deadliest Ones. https://www.strategy-business.com/blog/The-Upside-to-Robots-Stealing-Jobs-Theyll-Take-the-Deadliest-Ones?gko=9aea8
41. Matuzeviciute, K., Butkus, M., and Karaliute, A. (2017). Do technological innovations affect unemployment? Some empirical evidence from European countries. Economies 5(4), 1–19 (2017). https://www.mdpi.com/2227-7099/5/4/48.
42. Paul R. Krugman, 1994. Past and prospective causes of high unemployment. Economic Review, Federal Reserve Bank of Kansas City, issue Q IV, 23-43.
43. Pramudita, Y., Putro, S., Rochman, E., Rofiqi, A., Jauhari, A., Suzanti, I., & Rachmad, A. (2018). Clustering Sports News in Indonesian Using Modified K-Medoid Method. Proceedings of the The 1st International Conference on Computer Science and Engineering Technology Universitas Muria Kudus. doi:10.4108/eai.24-10-2018.2280569
44. RBC (2017). Humans Wanted: How Canadian youth can thrive in the age of disruption? http://www.rbc.com/newsroom/_assets-custom/pdf/03-2018-rbc-future-skills-report.pdf
45. Rotman, D. (2013). How Technology Is Destroying Jobs. MIT Technology Review. June 12, 2013. http://www.technologyreview.com/featuredstory/515926/how-technology-is-destroying-jobs/
46. Rousseeuw,P. J. (1987), Silhouettes: A graphical aid to the interpretation and validation of cluster analysis
47. Author links open overlay panelPeter J.Rousseeuw
48. Russell, S., and Norvig, P. (2010), Artificial Intelligence: A Modern Approach, 3rd edition, Pearson.
49. Smith, Adam. (2001). Wealth of Nations, edited by C. J. Bullock. Vol. X. The Harvard Classics. New York: P.F. Collier & Son, 1909–14; Bartleby.com, www.bartleby.com/10/.
50. Smith, M., Mikulecky, L., Kibby, M., Dreher, M., and Dole, J. (2000). RRQ Snippet: What Will Be the Demands of Literacy in the Workplace in the Next Millennium? Reading Research Quarterly, 35(3), 378-383. Retrieved from http://www.jstor.org/stable/748223
51. Spitz- Oener, A. (2006). Technical Change, Job Tasks, and Rising Educational Demands: Looking outside the Wage Structure. Journal of Labor Economics, 24(2), 235-270. doi:10.1086/499972
52. Strickland, E. (2016). Would You Trust a Robot Surgeon to Operate on You? Retrieved from https://spectrum.ieee.org/robotics/medical-robots/would-you-trust-a-robot-surgeon-to-operate-on-you
53. Takayama, L., and Nass, C. (2008). Beyond dirty, dangerous and dull: What everyday people think robots should do. HRI 2008 - Proceedings of the 3rd ACM/IEEE International Conference on Human-Robot Interaction: Living with Robots. 25-32. 10.1145/1349822.1349827.
54. Violante, G. L. (2016). Skill-biased Technical Change. The New Palgrave Dictionary of Economics, 2012 Version. doi:10.1057/9781137336583.1655

**Appendix: List of Vulnerable Occupations**

| Cluster | Occupations |
|---|---|
| 3 | Nursery and Greenhouse Managers |
| 3 | Farm and Ranch Managers |
| 3 | Licensing Examiners and Inspectors |
| 3 | Industrial Safety and Health Engineers |
| 3 | Nuclear Equipment Operation Technicians |
| 3 | Forest and Conservation Technicians |
| 3 | Occupational Health and Safety Specialists |
| 3 | Occupational Health and Safety Technicians |
| 3 | First-Line Supervisors of Correctional Officers |
| 3 | First-Line Supervisors of Police and Detectives |
| 3 | Municipal Fire Fighting and Prevention Supervisors |
| 3 | Forest Fire Fighting and Prevention Supervisors |
| 3 | Municipal Firefighters |
| 3 | Forest Firefighters |
| 3 | Immigration and Customs Inspectors |
| 3 | Fish and Game Wardens |
| 3 | Police Patrol Officers |
| 3 | Sheriffs and Deputy Sheriffs |
| 3 | Transit and Railroad Police |
| 3 | Animal Control Workers |
| 3 | First-Line Supervisors of Landscaping, |
| 3 | Landscaping and Groundskeeping Workers |
| 3 | Pesticide Handlers, Sprayers, and Applicators, Vegetation |
| 3 | Tree Trimmers and Pruners |
| 3 | Morticians, Undertakers, |
| 3 | Production, Planning, |
| 3 | First-Line Supervisors of Logging Workers |
| 3 | First-Line Supervisors of Aquacultural |
| 3 | First-Line Supervisors of Agricultural |
| 3 | Agricultural Equipment Operators |
| 3 | Farmworkers, Farm, Ranch, and Aquacultural Animals |
| 3 | Fallers |
| 3 | First-Line Supervisors of Construction |
| 3 | Boilermakers |
| 3 | Brickmasons and Blockmasons |
| 3 | Construction Carpenters |
| 3 | Rough Carpenters |
| 3 | Cement Masons and Concrete Finishers |
| 3 | Construction Laborers |
| 3 | Paving, Surfacing, and Tamping Equipment |
| 3 | Pile-Driver Operators |
| 3 | Operating Engineers and Other Construction |
| 3 | Electricians |
| 3 | Insulation Workers, Mechanical |
| 3 | Pipelayers |
| 3 | Pipe Fitters and Steamfitters |
| 3 | Plumbers |
| 3 | Reinforcing Iron and Rebar Workers |
| 3 | Roofers |
| 3 | Structural Iron and Steel Workers |
| 3 | Helpers--Brickmasons, Blockmasons, |
| 3 | Helpers--Carpenters |
| 3 | Elevator Installers and Repairers |
| 3 | Hazardous Materials Removal Workers |
| 3 | Highway Maintenance Workers |
| 3 | Rail-Track Laying |
| 3 | Septic Tank Servicers and Sewer Pipe Cleaners |
| 3 | Weatherization Installers and Technicians |
| 3 | Derrick Operators, Oil and Gas |
| 3 | Rotary Drill Operators, Oil and Gas |
| 3 | Service Unit Operators, Oil, Gas, and Mining |
| 3 | Earth Drillers, Except Oil and Gas |
| 3 | Explosives Workers, Ordnance Handling Experts |
| 3 | Continuous Mining Machine Operators |
| 3 | Mine Cutting and Channeling Machine Operators |
| 3 | Rock Splitters, Quarry |
| 3 | Roof Bolters, Mining |
| 3 | Roustabouts, Oil and Gas |
| 3 | Helpers--Extraction Workers |
| 3 | First-Line Supervisors of Mechanics, Installers |
| 3 | Electric Motor, Power Tool, |
| 3 | Electrical and Electronics Repairers, Powerhouse, |
| 3 | Aircraft Mechanics and Service Technicians |
| 3 | Automotive Master Mechanics |
| 3 | Automotive Specialty Technicians |
| 3 | Bus and Truck Mechanics |
| 3 | Farm Equipment Mechanics |
| 3 | Mobile Heavy Equipment Mechanics |
| 3 | Motorboat Mechanics and Service Technicians |
| 3 | Outdoor Power Equipment and Other Mechanics |
| 3 | Recreational Vehicle Service Technicians |
| 3 | Tire Repairers and Changers |
| 3 | Control and Valve Installers and Repairers, Door |
| 3 | Heating and Air Conditioning Mechanics |
| 3 | Refrigeration Mechanics and Installers |
| 3 | Industrial Machinery Mechanics |
| 3 | Maintenance Workers, Machinery |

| | | | |
|---|---|---|---|
| 3 | Millwrights | 3 | Cutting and Slicing Machine Setters, |
| 3 | Refractory Materials Repairers, Except | 3 | Furnace, Kiln, Oven, Drier, and Kettle |
| 3 | Electrical Power-Line Installers and Repairers | 3 | Inspectors, Testers, Sorters, Samplers |
| 3 | Telecommunications Line Installers and Repairers | 3 | Segmental Pavers |
| 3 | Maintenance and Repair Workers, General | 3 | Medical Appliance Technicians |
| 3 | Wind Turbine Service Technicians | 3 | Cooling and Freezing Equipment |
| 3 | Commercial Divers | 3 | Glass Blowers, Molders, Benders, and Finishers |
| 3 | Manufactured Building and Installers | 3 | Aircraft Cargo Handling Supervisors |
| 3 | Riggers | 3 | First-Line Supervisors of Helpers Movers, Hand |
| 3 | Signal and Track Switch Repairers | 3 | Recycling Coordinators |
| 3 | Geothermal Technicians | 3 | Airline Pilots, Copilots, and Flight Engineers |
| 3 | Fiberglass Laminators and Fabricators | 3 | Commercial Pilots |
| 3 | Rolling Machine Setters, Operators | 3 | Airfield Operations Specialists |
| 3 | Cutting, Punching, and Press Machine | 3 | Locomotive Engineers |
| 3 | Metal-Refining Furnace Operators | 3 | Locomotive Firers |
| 3 | Pourers and Casters, Metal | 3 | Rail Yard Engineers, Dinkey Operators, |
| 3 | Model Makers, Metal and Plastic | 3 | Railroad Brake, Signal, and Switch Operators |
| 3 | Patternmakers, Metal and Plastic | 3 | Railroad Conductors and Yardmasters |
| 3 | Molding, Coremaking, and Casting | 3 | Sailors and Marine Oilers |
| 3 | Multiple Machine Tool Setters, | 3 | Ship and Boat Captains |
| 3 | Tool and Die Makers | 3 | Mates- Ship, Boat, and Barge |
| 3 | Heat Treating Equipment Setters, | 3 | Pilots, Ship |
| 3 | Layout Workers, Metal and Plastic | 3 | Motorboat Operators |
| 3 | Plating and Coating Machine Setters | 3 | Ship Engineers |
| 3 | Printing Press Operators | 3 | Automotive and Watercraft Service Attendants |
| 3 | Extruding and Forming Machine Setters, | 3 | Crane and Tower Operators |
| 3 | Model Makers, Wood | 3 | Excavating and Loading Machine and Dragline |
| 3 | Patternmakers, Wood | 3 | Loading Machine Operators, Underground |
| 3 | Sawing Machine Setters, Operators, | 3 | Hoist and Winch Operators |
| 3 | Woodworking Machine Setters, Operators, and Tenders, Except Sawing | 3 | Industrial Truck and Tractor Operators |
| 3 | Power Plant Operators | 3 | Gas Compressor and Gas Pumping Station |
| 3 | Stationary Engineers and Boiler Operators | 3 | Pump Operators, Except Wellhead Pumpers |
| 3 | Water and Wastewater Treatment Plant | 3 | Wellhead Pumpers |
| 3 | Chemical Plant and System Operators | 3 | Refuse and Recyclable Material Collectors |
| 3 | Gas Plant Operators | 3 | Mine Shuttle Car Operators |
| 3 | Petroleum Pump System Operators, Refinery Operators, and Gaugers | 3 | Tank Car, Truck, and Ship Loaders |
| 3 | Biofuels Processing Technicians | 3 | Dredge Operators |
| 3 | Biomass Plant Technicians | 3 | Cleaners of Vehicles and Equipment |
| 3 | Hydroelectric Plant Technicians | 3 | Machine Feeders and Offbearers |
| 3 | Chemical Equipment Operators | 3 | Fishers and Related Fishing Workers |
| 3 | Separating, Filtering, Clarifying, | 3 | Hunters and Trappers |
| 3 | Crushing, Grinding, and Polishing Machine Setters | 3 | Forest and Conservation Workers |
| 3 | Mixing and Blending Machine Setters, Tenders | 3 | Log Graders and Scalers |
| 3 | Extruding, Forming, Pressing, Setters, Tenders | 3 | Logging Equipment Operators |
| 3 | Cutters and Trimmers, Hand | 6 | Loan Counselors |
| | | 6 | Computer Network Support Specialists |

| 6 | Electrical Drafters | 6 | Telemarketers |
|---|---|---|---|
| 6 | Paralegals and Legal Assistants | 6 | Switchboard Operators |
| 6 | Court Reporters | 6 | Telephone Operators |
| 6 | Title Examiners, Abstractors, and Searchers | 6 | Statement Clerks |
| 6 | Library Technicians | 6 | Billing, Cost, and Rate Clerks |
| 6 | Technical Writers | 6 | Bookkeeping, Accounting, and Auditing Clerks |
| 6 | Radio Operators | 6 | Gaming Cage Workers |
| 6 | Pharmacy Technicians | 6 | Payroll and Timekeeping Clerks |
| 6 | Ophthalmic Medical Technicians | 6 | Procurement Clerks |
| 6 | Medical Records and Health Information | 6 | Tellers |
| 6 | Ophthalmic Medical Technologists | 6 | Brokerage Clerks |
| 6 | Orderlies | 6 | Correspondence Clerks |
| 6 | Medical Transcriptionists | 6 | Court Clerks |
| 6 | Pharmacy Aides | 6 | Municipal Clerks |
| 6 | Bailiffs | 6 | License Clerks |
| 6 | Transportation Security Screeners | 6 | Credit Authorizers |
| 6 | Cooks, Fast Food | 6 | Credit Checkers |
| 6 | Cooks, Short Order | 6 | Customer Service Representatives |
| 6 | Bartenders | 6 | Eligibility Interviewers, Government Programs |
| 6 | Combined Food Preparation and Serving | 6 | File Clerks |
| 6 | Counter Attendants, Cafeteria, Food Concession, and Coffee Shop | 6 | Hotel, Motel, and Resort Desk Clerks |
| 6 | Baristas | 6 | Interviewers, Except Eligibility and Loan |
| 6 | Waiters and Waitresses | 6 | Library Assistants, Clerical |
| 6 | Food Servers, Nonrestaurant | 6 | Order Clerks |
| 6 | Dining Room and Cafeteria Attendants and Bartender Helpers | 6 | Human Resources Assistants, |
| 6 | Hosts and Hostesses, Restaurant, Lounge, and Coffee Shop | 6 | Receptionists and Information Clerks |
| 6 | Gaming Supervisors | 6 | Reservation and Transportation Ticket |
| 6 | Slot Supervisors | 6 | Couriers and Messengers |
| 6 | Gaming Dealers | 6 | Police, Fire, and Ambulance Dispatchers |
| 6 | Gaming and Sports Book Writers and Runners | 6 | Postal Service Clerks |
| 6 | Ushers, Lobby Attendants, and Ticket Takers | 6 | Postal Service Mail Carriers |
| 6 | Amusement and Recreation Attendants | 6 | Postal Service Mail Sorters, Processors |
| 6 | Locker Room, Coatroom, and Dressing Room Attendants | 6 | Stock Clerks, Sales Floor |
| 6 | Barbers | 6 | Marking Clerks |
| 6 | Manicurists and Pedicurists | 6 | Stock Clerks- Stockroom, Warehouse, |
| 6 | Shampooers | 6 | Order Fillers, Wholesale and Retail Sales |
| 6 | Skincare Specialists | 6 | Executive Secretaries and Executive Administrative Assistants |
| 6 | First-Line Supervisors of Retail Sales Workers | 6 | Legal Secretaries |
| 6 | Cashiers | 6 | Medical Secretaries |
| 6 | Gaming Change Persons and Booth Cashiers | 6 | Secretaries and Administrative Assistants, Except Legal, Medical, and Executive |
| 6 | Counter and Rental Clerks | 6 | Data Entry Keyers |
| 6 | Demonstrators and Product Promoters | 6 | Word Processors and Typists |
| 6 | Models | 6 | Insurance Claims Clerks |
|   |   | 6 | Insurance Policy Processing Clerks |
| 6 | Mail Clerks and Mail Machine Operators, Except Postal Service | 7 | Funeral Attendants |
| 6 | Office Clerks, General | 7 | Baggage Porters and Bellhops |
| 6 | Proofreaders and Copy Markers | 7 | Meter Readers, Utilities |
| 6 | Graders and Sorters, Agricultural Products | 7 | Shipping, Receiving, and Traffic Clerks |
| 6 | Nursery Workers | 7 | Weighers, Measurers, Checkers, |
| 6 | Slaughterers and Meat Packers | 7 | Office Machine Operators, Except Computer |
| 6 | Laundry and Dry-Cleaning Workers | 7 | First-Line Supervisors of Animal Husbandry |
| 6 | Ophthalmic Laboratory Technicians | 7 | Agricultural Inspectors |
| 6 | Laborers and Freight, Stock | 7 | Animal Breeders |
| 7 | Environmental Compliance Inspectors | 7 | Farmworkers and Laborers, Crop |
| 7 | Farm Labor Contractors | 7 | Fishers and Related Fishing Workers |
| 7 | Energy Auditors | 7 | Hunters and Trappers |
| 7 | Environmental Engineering Technicians | 7 | Forest and Conservation Workers |
| 7 | Non-Destructive Testing Specialists | 7 | Logging Equipment Operators |
| 7 | Electrical Engineering Technologists | 7 | Log Graders and Scalers |
| 7 | Industrial Engineering Technologists | 7 | Stonemasons |
| 7 | Manufacturing Production Technicians | 7 | Carpet Installers |
| 7 | Surveying Technicians | 7 | Floor Layers, Except Carpet, Wood, |
| 7 | Agricultural Technicians | 7 | Floor Sanders and Finishers |
| 7 | Food Science Technicians | 7 | Tile and Marble Setters |
| 7 | Environmental Science and Protection Technicians, Including Health | 7 | Terrazzo Workers and Finishers |
| 7 | Athletes and Sports Competitors | 7 | Drywall and Ceiling Tile Installers |
| 7 | Histotechnologists and Histologic Technicians | 7 | Tapers |
| 7 | Dietetic Technicians | 7 | Glaziers |
| 7 | Home Health Aides | 7 | Insulation Workers, Floor, Ceiling, and Wall |
| 7 | Fire Inspectors | 7 | Painters, Construction and Maintenance |
| 7 | Police Identification and Records Officers | 7 | Paperhangers |
| 7 | Parking Enforcement Workers | 7 | Plasterers and Stucco Masons |
| 7 | Gaming Surveillance Officers and Gaming Investigators | 7 | Sheet Metal Workers |
| 7 | Security Guards | 7 | Solar Photovoltaic Installers |
| 7 | Crossing Guards | 7 | Helpers--Electricians |
| 7 | First-Line Supervisors of Food Preparation and Serving Workers | 7 | Helpers--Painters, Paperhangers, Plasterers |
| 7 | Cooks, Institution and Cafeteria | 7 | Helpers--Pipelayers, Plumbers, Pipefitters |
| 7 | Food Preparation Workers | 7 | Helpers--Roofers |
| 7 | Dishwashers | 7 | Construction and Building Inspectors |
| 7 | First-Line Supervisors of Housekeeping and Janitorial Workers | 7 | Fence Erectors |
| 7 | Janitors and Cleaners, Except Maids and Housekeeping Cleaners | 7 | Segmental Pavers |
| 7 | Maids and Housekeeping Cleaners | 7 | Solar Thermal Installers and Technicians |
| 7 | Pest Control Workers | 7 | Radio, Cellular, and Tower Equipment |
| 7 | Animal Trainers | 7 | Radio Mechanics |
| 7 | Nonfarm Animal Caretakers | 7 | Telecommunications Equipment Installers and Repairers, Except Line Installers |
| 7 | Motion Picture Projectionists | 7 | Avionics Technicians |
| | | 7 | Electrical and Electronics Installers and Repairers, Transportation Equipment |
| | | 7 | Electrical and Electronics Repairers, Commercial and Industrial Equipment |
| 7 | Electronic Equipment Installers and Repairers, Motor Vehicles | 7 | Shoe Machine Operators and Tenders |
| 7 | Security and Fire Alarm Systems Installers | 7 | Upholsterers |
| 7 | Automotive Body and Related Repairers | 7 | Cabinetmakers and Bench Carpenters |
| 7 | Automotive Glass Installers and Repairers | 7 | Furniture Finishers |
| 7 | Rail Car Repairers | 7 | Grinding and Polishing Workers, Hand |
| 7 | Motorcycle Mechanics | 7 | Cutters and Trimmers, Hand |
| 7 | Bicycle Repairers | 7 | Gem and Diamond Workers |
| 7 | Mechanical Door Repairers | 7 | Precious Metal Workers |
| 7 | Home Appliance Repairers | 7 | Dental Laboratory Technicians |
| 7 | Medical Equipment Repairers | 7 | Etchers and Engravers |
| 7 | Musical Instrument Repairers and Tuners | 7 | Stone Cutters and Carvers, Manufacturing |
| 7 | Watch Repairers | 7 | Molding and Casting Workers |
| 7 | Coin, Vending, and Amusement Machine Servicers and Repairers | 7 | Paper Goods Machine Setters, Operators, |
| | | 7 | Tire Builders |
| 7 | Fabric Menders, Except Garment | 7 | Helpers--Production Workers |
| 7 | Locksmiths and Safe Repairers | 7 | Recycling and Reclamation Workers |
| 7 | Electromechanical Equipment Assemblers | 7 | Ambulance Drivers and Attendants, |
| 7 | Engine and Other Machine Assemblers | 7 | Bus Drivers, Transit and Intercity |
| 7 | Structural Metal Fabricators and Fitters | 7 | Bus Drivers, School or Special Client |
| 7 | Team Assemblers | 7 | Driver/Sales Workers |
| 7 | Timing Device Assemblers and Adjusters | 7 | Heavy and Tractor-Trailer Truck Drivers |
| 7 | Bakers | 7 | Light Truck or Delivery Services Drivers |
| 7 | Butchers and Meat Cutters | 7 | Taxi Drivers and Chauffeurs |
| 7 | Meat, Poultry, and Fish Cutters and Trimmers | 7 | Subway and Streetcar Operators |
| 7 | Food and Tobacco Roasting, Baking | 7 | Bridge and Lock Tenders |
| 7 | Food Batchmakers | 7 | Parking Lot Attendants |
| 7 | Food Cooking Machine Operators | 7 | Traffic Technicians |
| 7 | Computer-Controlled Machine Tool | 7 | Aviation Inspectors |
| 7 | Computer Numerically Controlled | 7 | Transportation Vehicle, Equipment Inspectors, |
| 7 | Extruding and Drawing Machine | 7 | Freight and Cargo Inspectors |
| 7 | Forging Machine Setters, Operators, | 7 | Conveyor Operators and Tenders |
| 7 | Drilling and Boring Machine Tool Setters | 7 | Dredge Operators |
| 7 | Grinding, Lapping, Polishing, | 7 | Cleaners of Vehicles and Equipment |
| 7 | Lathe and Turning Machine Tool Setters, | 7 | Machine Feeders and Offbearers |
| 7 | Milling and Planing Machine Setters, | 7 | Packers and Packagers, Hand |
| 7 | Machinists | | |
| 7 | Foundry Mold and Coremakers | | |
| 7 | Welders, Cutters, and Welder Fitters | | |
| 7 | Solderers and Brazers | | |
| 7 | Welding, Soldering, and Brazing Machine | | |
| 7 | Tool Grinders, Filers, and Sharpeners | | |
| 7 | Prepress Technicians and Workers | | |
| 7 | Print Binding and Finishing Workers | | |
| 7 | Pressers, Textile, Garment, and Related Materials | | |
| 7 | Sewing Machine Operators | | |
| 7 | Shoe and Leather Workers and Repairers | | |